# Calculation of extended gcd by normalization

-------------------------------------------------

Number theory

-------------------------------------------------

July 7, 2018

**WOLF Marc, WOLF François, LE COZ Corentin**
Independent researchers.
Email: marc.wolf3@wanadoo.fr
Email: francois.wolf@dbmail.com
Email : corentin0107@hotmail.fr

#### Abstract

We propose a new algorithm solving the extended gcd problem, which provides a solution minimizing one of the two coordinates. The algorithm relies on elementary arithmetic properties.

# Contents

# 1    Introduction

## 1.1    The problem of extended gcd

Let $a$ and $b$ be two integers, the extended gcd problem consists in finding three integers $u$, $v$ and $g$ such that:

$$ua + vb = g$$

with $g$ equal to gcd of $a$ and $b$.

In number theory, this problem occurs in many situations like in the theory of corrector codes [1] or for the factorization of integers [2]. Moreover, it is the elementary component of the classical algorithm of Smith invariants computation of a matrix with integer coefficients, which allows the general resolution of the linear diophantine equations [3]. See also [4] for a recent application of the Euclid algorithm.

## 1.2    Notation

Let $x$ and $y$ be two integers, we denote $x \bmod y$ the remainder of the Euclidean division of $x$ by $y$, which belongs to $[\![0, y-1]\!]$ by convention.

# 2    Extended GCD with $a$ and $b$ coprime integers.

The purpose of this section is to propose an algorithm to solve the extended gcd problem in the case of *coprime* integers. Let us take $a, b \in \mathbb{N}^*$.

## 2.1    The normalizer $v_c$

For any integer $c$, we consider the following equation $(E_c)$ in $u$ and $v$:

$$ua + vb = c \qquad\qquad (E_c)$$

It is well-known that solutions exist if and only if $\gcd(a, b) \,|\, c$, and that given a particular solution $(u_0, v_0)$, the set of all solutions is equal to $\{(u_0 + kb, v_0 - ka), k \in \mathbb{Z}\}$.

**Definition 1.** *Assume* $\gcd(a, b) \,|\, c$. *Then there exists a unique solution* $(u_c, v_c)$ *of* $(E_c)$ *verifying* $v_c \in [\![0, a-1]\!]$, *which will be called the* normal *solution.* $v_c$ *will be called the* normalizer *of* $(E_c)$.

**Remark.** *Knowing* $v_c$, *we can deduce* $u_c = (c - v_c b)/a$, *and therefore all the solutions of* $(E_c)$. *Furthermore, if* $c \in [\![0, ab-1]\!]$, $u_c \in [\![-b+1, b-1]\!]$, *and if* $c = 1$ *and* $b > 1$, $u_c \in [\![-b+1, -1]\!]$.

From the structure of solutions of $(E_c)$, we get the following proposition:

**Proposition 1.** *For any solution* $(u, v)$ *of* $(E_c)$, $v \equiv v_c \ (mod\ a)$.

**Definition 2.** *Under the same assumptions as definition 1, we can define the co-normalizer of* $(E_c)$ *to be the unique* $t_c \in [\![0, a-1]\!]$ *such that* $(x_c, -t_c)$ *solves* $(E_c)$ *for some* $x_c$.

## 2.2 Arithmetic properties of the normalizer

Let $g = \gcd(a, b)$.

**Proposition 2.** *The normalizer is additive:*
$$\forall c, c' \in \mathbb{Z}g, \ v_{c+c'} \equiv v_c + v_{c'} \pmod{a}$$
It is easily proved by summing the equations $(E_c)$ and $(E_{c'})$.

**Corollary 1.** *The normalizer is multiplicative:*
$$\forall (x, c) \in \mathbb{Z} \times \mathbb{Z}g, v_{xc} \equiv x.v_c \pmod{a}.$$

*In particular:*
$$\forall c \in \mathbb{Z}, v_{cg} \equiv c.v_g \pmod{a}.$$

From the previous propositions, we deduce that the normalizer is $\mathbb{Z}$-*linear* in $c$ modulo $a$, but for our algorithm we will only need the following corollary:

**Corollary 2.** *We have:*
$$\forall c, c' \in \mathbb{Z}g, v_{c-c'} \equiv v_c - v_{c'} \pmod{a}.$$

Under certain assumptions, the normalizer may also be stable by division:

**Proposition 3.** *If* $a$ *is odd and* $c \in \mathbb{Z}g$ *is even, then* $\frac{c}{2} \in \mathbb{Z}g$ *and:*
$$v_{c/2} = \begin{cases} \dfrac{v_c}{2} & \text{if } v_c \text{ is even,} \\ \dfrac{v_c + a}{2} & \text{otherwise.} \end{cases}$$

*Proof.* Assume $a$ odd, $c$ even, $u_c a + v_c b = c$. The fact that $\frac{c}{2} \in \mathbb{Z}g$ will follow from the fact that we will write it as a linear combination of $a$ and $b$.

- If $v_c$ is even, $u_c a = c - v_c b$ is even as well, and so is $u_c$. Therefore $\frac{u_c}{2}, \frac{v_c}{2}, \frac{c}{2} \in \mathbb{Z}$ verify $\frac{u_c}{2} a + \frac{v_c}{2} b = \frac{c}{2}$, which clearly implies $v_{c/2} = \frac{v_c}{2}$.
- If $v_c$ is odd:
    - Either $u_c$ is odd, which means that $v_c b = c - u_c a$ is odd, so $b$ must be odd.
    - Or $u_c$ is even, in which case $v_c b = c - u_c a$ is even and so $b$ must be even too.

In both cases, $\frac{u_c - b}{2}, \frac{v_c + a}{2}, \frac{c}{2} \in \mathbb{Z}$ still verify $\frac{u_c - b}{2} a + \frac{v_c + a}{2} b = \frac{c}{2}$, and $0 < \frac{v_c + a}{2} \leq a - \frac{1}{2}$, from which we deduce that $v_{c/2} = \frac{v_c + a}{2}$.

**Remark.** *Similar properties for the co-normalizer can be also shown.*

We deduce from the above proposition the following algorithm, which will be part of the extended gcd algorithm: given $(a, c, v)$, such that $a$ is odd, and $v$ is the normalizer of some $(E_c)$, it returns the pair of similar integers $(c', v')$ obtained by dividing $c$ by $2$ as many times as possible.

---

**Algorithm 1: Function Div1(a,c,v)**

---

**While** $c \bmod 2 = 0$ **do**
  $c \leftarrow c/2$
  **If** $v$ is even **then**
    $v \leftarrow v/2$
  **Else**
    $v \leftarrow (v+a)/2$
  **End If**
**End While**         $\rightarrow$ we divide as many times as possible $c$ by 2.
**Return** $(c, v)$

---

We determine two particular values of the normalizer that will initiate our algorithm:
**Proposition 4.** *The normalizer verifies:*
 • if $c = b \bmod a$, $v_c = 1$
 • if $c = -b \bmod a$, $v_c = a - 1$

*Proof.*
**Case 1: $c = b \bmod a$**
Let $q$ be the quotient of the Euclidean division of $b$ by $a$, i.e. $b = aq + c$. Hence comes the following equality:
$$-qa + b = c$$
i.e. $(-q, 1)$ is the normal solution of $(E_c)$ and $v_c = 1$.
 **Case 2: $c = -b \bmod a$**
 We know that $v_{-c} = 1$, therefore $v_c = -1 \bmod a = a - 1$.

## 2.3  Presentation of the algorithm

 Solving the problem of extended gcd with $a$ and $b$ being coprime integers is equivalent to determining $v_1$. The algorithm **WWL1** proceeds in two steps: an initialization with the two values of $v_c$ given by the previous proposition, and then a descent loop using corollary 2 of proposition 2 to determine $v_1$.

 Once $v_1$ has been determined, the complete solution is obtained by calculating:
$$u_1 = \frac{1 - v_1 b}{a}.$$
Then $(u_1, v_1)$ is a solution of $(E_1)$. Because the Euclidean division of $b$ by $a$ is done at the beginning of the algorithm, say $b = aq + r$, we put directly $u_1 = -v_1 q + \frac{1 - v_1 r}{a}$, which can be advantageous if $b$ is large.

---

**Algorithm 2: WWL1(a,b)** Given $a$ and $b$ two integers, with $a$ and $b$ coprimes and $a$ odd,

---

returns the unique pair of integers $(u, v)$ such that $ua + vb = 1$ and $v \in [\![0, a-1]\!]$

---

***First step: initialization.***

    $c_1 \leftarrow b \bmod a, \ c_2 \leftarrow a - c_1$

    $v_1 \leftarrow 1, \ v_2 \leftarrow a - 1$         $\rightarrow$ initialization of the two values of normalizer $v_c$.

    $(c_1, v_1) \leftarrow \textbf{Div1}(a, c_1, v_1)$   $\rightarrow$ we divide as many times as possible $c_1$ by 2.

    $(c_2, v_2) \leftarrow \textbf{Div1}(a, c_2, v_2)$   $\rightarrow$ we divide as many times as possible $c_2$ by 2.

    **If** $c_2 < c_1$ **then**

        $(c_1, c_2) \leftarrow (c_2, c_1), \ (v_1, v_2) \leftarrow (v_2, v_1)$

    **End If**                 $\rightarrow$ we ensure that $c_1 < c_2$

***Second step: iteration.***

    **While** $c_1 > 1$ **do**

        $c_2 \leftarrow c_2 - c_1$

        **If** $v_2 - v_1 < 0$ **then**

            $v_2 \leftarrow v_2 - v_1 + a$

        **Else**

            $v_2 \leftarrow v_2 - v_1$

        **End If**     $\rightarrow$ we modify $c_2$ and we compute the associated normalizer $v_2$.

        $(c_2, v_2) \leftarrow \textbf{Div1}(a, c_2, v_2)$ $\rightarrow$ we divide $c_2$ by 2 as many times as possible.

        **If** $c_2 < c_1$ **then**

            $(c_1, c_2) \leftarrow (c_2, c_1), \ (v_1, v_2) \leftarrow (v_2, v_1)$

        **End If**         $\rightarrow$ we reassign $c_1$ and $c_2$ so as to have $c_1 < c_2$.

    **End While**         $\rightarrow$ we will leave the loop when $c_1 = 1$

    $v \leftarrow v_1$

    $u_1 \leftarrow -vq, \ u_2 \leftarrow (1 - vr)/a$

    $u \leftarrow u_1 - u_2$

    **Return** $(u, v)$         $\rightarrow$ the solution is returned

---

## 2.4   Validity of the algorithm

    It can be noted that throughout the loop, we keep $v_i$ equal to the normalizer of $c_i$. Furthermore, at each step of the loop, $c_1 + c_2$ decreases, with $c_1$ and $c_2$ remaining positive. This proves the *termination* of the algorithm (the sequence of $c_1 + c_2$ has to be <u>finite</u>).

    We deduce that at the end of the loop $c_1 = 0$ or 1. However, because $\gcd(c_1, c_2) = \gcd(c_1, c_2 - c_1)$, we deduce that $\gcd(c_1, c_2)$ remains constant equal to its initial value $\gcd(b \bmod a, a - (b \bmod a)) = \gcd(a, b) = 1$ by hypothesis. Hence at the end of the loop, if $c_1$ was equal to 0, this would necessarily mean than $c_2 = 1$, which is impossible: otherwise at the previous step we would already have one of the $c_i \le 1$, which would mean that we should already have exited the loop.

---

# 3 General case

It is impractical to impose for $a$ and $b$ to be coprimes, therefore we will adapt the previous algorithm to the general case where $a$ and $b$ are any numbers. Experimental tests also highlighted that it is more advantageous to calculate $u_c$ at the same time as the normalizer $v_c$ during the finite descent. Hence our algorithm will return, when it is possible, the unique triplet $(u, v, g)$ satisfying both conditions $ua + vb = g = \gcd(a, b)$ and $v \in [\![0, a-1]\!]$.

If $a$ and $b$ are both even, a first step is to factor them both by the greatest power of $2$ possible (which boils down to a cheap bit shift), so that one of them necessarily becomes odd. From now on, let us assume by symmetry that $a$ is *odd*.

---

**Algorithm 3: Function Div2(u,v,c)**: Given $(a, b, u, v, c)$ integers verifying the conditions $a = 1 \bmod 2$, $ua + vb = c$ and $v \in [\![0, a-1]\!]$, returns a triplet $(u', v', c')$ verifying the same conditions, obtained by dividing as many times as possible $c$ by $2$.

---

**While** $c = 0 \bmod 2$ **do**

    **If** $v = 0 \bmod 2$ **then**

$$(u, v, c) \leftarrow \left(\frac{u}{2}, \frac{v}{2}, \frac{c}{2}\right)$$

    **Else**

$$(u, v, c) \leftarrow \left(\frac{u-b}{2}, \frac{v+a}{2}, \frac{c}{2}\right)$$

    **End If**

**End While**

**Return** $(u, v, c)$

---

*Proof.* This is an extension of the algorithm **Div1**, exploiting directly the result of the second corollary of proposition 2.

.

---

**Algorithm 4: WWL2**: Given $a$ and $b$ be two integers, with $a$ odd, returns a triplet $(u, v, g)$ such that $g =$pgcd$(a, b)$, $ua + vb = g$ and $v \in [\![0, a-1]\!]$.

---

***First step: initialization.***

    $c_1 \leftarrow b - \left\lfloor \frac{b}{a} \right\rfloor a$, $c_2 \leftarrow a - c_1$, $v_1 \leftarrow 1$, $v_2 \leftarrow a - 1$

    $u_1 \leftarrow \frac{c_1 - v_1 b}{a}$, $u_2 \leftarrow 1 - u_1 - b$       → initialization of two triplets $(u, v, c)$

    $(u_1, v_1, c_1) \leftarrow Div2(u_1, v_1, c_1)$    → we divide as many times as possible $c_1$ by 2.

    $(u_2, v_2, c_2) \leftarrow Div2(u_2, v_2, c_2)$    → we divide as many times as possible $c_2$ by 2.

    **If** $c_2 < c_1$ **then**

        $(u_1, v_1, c_1, u_2, v_2, c_2) \leftarrow (u_2, v_2, c_2, u_1, v_1, c_1)$

    **End If**                → we ensure that $c_1 < c_2$

***Second step: iteration.***

    **While** $c_1 > 0$ **do**

---

$$c_2 \leftarrow c_2 - c_1$$

**If** $v_2 - v_1 < 0$ **then**

$$v_2 \leftarrow v_2 + a - v_1, \ u_2 \leftarrow u_2 - u_1 - b$$

**Else**

$$v_2 \leftarrow v_2 - v_1, \ u_2 \leftarrow u_2 - u_1$$

**End If** $\qquad \rightarrow$ we modify $u_2$ and $v_2$ so that they verify $(E_{c_2})$.

$$(c_2, v_2) \leftarrow Div2(c_2, v_2) \qquad \rightarrow$$ we divide as many times as possible $c_2$ by 2.

**If** $c_2 < c_1$ **then**

$$(u_1, v_1, c_1, u_2, v_2, c_2) \leftarrow (u_2, v_2, c_2, u_1, v_1, c_1)$$

**End If** $\qquad \rightarrow$ we reassign $c_1$ and $c_2$ so as to have $c_1 < c_2$.

$$c_2 \leftarrow c_2 - c_1$$

**End While** $\qquad \rightarrow$ the loop is left when $c_1 = 0$

**Return** $(u_2, v_2, c_2)$ $\qquad \rightarrow$ the solution is returned

---

*Proof.* The same arguments as in the proof of **WWL1** hold: $c_1 + c_2$ decreases strictly along the loop, which ensures that it terminates, $(u_i, v_i)$ always remains the normal solution of $(E_{c_i})$ along the loop, and $\gcd(c_1, c_2)$ stays equal to $\gcd(a, b)$, so that when $c_1 = 0$ necessarily $c_2 = \gcd(a, b)$.

**Remark 1.** Here the exit condition is somehow simpler but it actually means one more step than **WWL1** when $\gcd(a, b) = 1$, which could be optimized away in the coprime case.

**Remark 2.** *The algorithm can be adapted to return the co-normalizer instead of the normalizer. This is left to the reader.*

# 4    Conclusion

We have reduced linear Diophantine equations solutions to a single and unique integer, its normalizer. We have used this terminology to develop a solution of the extended gcd problem, which has the advantage of controlling the size of the final result as well as that of the intermediary steps, while remaining very simple to implement. Our benchmark with other algorithms (see appendix) suggested a similar complexity, but an average computational gain of 15% which makes it advantageous to use. Optimizing the algorithm was not our main concern for this article, however we discuss some further possible improvements at the end of the appendix.

Acknowledgments: We would like to thank François-Xavier Villemin for his attentive comments and suggestions.

# 6 Appendix: gcd algorithms

We present here two popular gcd algorithms (not in their extended version for the sake of simplicity), namely the Euclidean algorithm [5] and its binary version [6]. We note that the steps followed by our **WWL2** algorithm are exactly the same as a combination of those two, and as stated in the conclusion of our article outperforms on average those two.

---

**Algorithm 5: (classical Euclid algorithm) EulerGCD:** *Given $a, b$ two integers, returns their gcd.*

---

$r_1, r_2 \leftarrow a, b$
**While** $r_1 > 0$ **do**
$\qquad r_1, r_2 \leftarrow r_2 \bmod r_1, r_1$
**End While**
**Return** $r_2$

---

This algorithm is the most self-contained, however it performs quite badly due to the high numerical cost of the Euclidean divisions. The next one takes advantage of the inexpensivity of divisions by 2 (bit shifts):

---

**Algorithm 6: (binary Euclid algorithm) BinaryGCD:** *Given $a, b$ two integers, returns their gcd.*

---

$m = 0$
**While** $a, b = 0 \bmod 2$ **do**
$\qquad m, a, b \leftarrow m + 1, \frac{a}{2}, \frac{b}{2}$
**End While** $\qquad\qquad\qquad\qquad \rightarrow$ Factorization of a power of $2$
$r_1, r_2 \leftarrow \min(a, b), \max(a, b)$
**While** $r_1 > 0$ **do**
$\qquad r_2 \leftarrow r_2 - r_1$
$\qquad$**While** $r_2 = 0 \bmod 2$ **do**
$\qquad\qquad r_2 \leftarrow r_2/2$
$\qquad$**End While**
$\qquad r_1, r_2 \leftarrow \min(r_1, r_2), \max(r_1, r_2)$

---

**End While**
**Return** $r_2 \times 2^m$

---

As stated before, the following mixed Euclid algorithm is a mixture of the two previous ones, which only keeps a single Euclidean division:

---

**Algorithm 7**: (**mixed Euclid algorithm**) **MixedEuclid:** *Given* $a, b$ *two integers, returns their gcd.*

---

$m = 0$
**While** $a$ and $b$ are even **do**
    $m, a, b \leftarrow m + 1, \frac{a}{2}, \frac{b}{2}$
**End While**                  → Computation of the common power of $2$
$r_1, r_2 \leftarrow \min(a, b), \max(a, b)$
$(r1, r2) \leftarrow (r2 \bmod r1, r1)$        → Single Euclidean division
**Return** $BinaryGCD(r_1, r_2)$       →We apply the binary algorithm

---

On average, the cost of this single division is compensated by a significant decrease in the number of remaining binary algorithm steps. For large integers, it could be interesting to check whether performing more than one Euclidean division yields yet a more significant gain. Another idea would be to trigger an Euclidean division when the binary algorithm is likely to perform many substractions in a row (typically if $r_1$ is large and odd, and $r_2$ is small and even).